\documentclass[12pt,a4paper]{article}
\usepackage{amsfonts,latexsym}
\usepackage{amsmath,amssymb}
\usepackage{graphicx,color}

\renewcommand{\thefootnote}{\fnsymbol{footnote}}

\begin{document}

\vspace{12mm}

\begin{center}
{{{\Large {\bf Thermodynamic analysis and shadow bound of black holes surrounded by a dark matter halo }}}}\\[10mm]

{Yun Soo Myung\footnote{e-mail address: ysmyung@inje.ac.kr}}\\[8mm]

{Center for Quantum Spacetime, Sogang University, Seoul 04107, Republic of  Korea\\[0pt] }

\end{center}
\vspace{2mm}

\begin{abstract}
We perform the thermodynamic analysis of a black hole (BH) immersed in a dark matter halo (DMH). It is shown that  the BH could  not be in thermal equilibrium with the DMH in
any regions outside the event horizon.  This means that  the thermodynamic influence of the environment (DMH) is relatively small on the BH.
Importantly,   it  does not alter the nature of the negative heat capacity for the BH.
We stress that the Newtonian ($1/a_0$) approximation gives us  a correct thermodynamic description for the BH surrounded by DMH  because the first law of thermodynamics and Smarr formula are satisfied.
Hence, the Newtonian Helmholtz free energy is employed  to reveal  that there is the absence of  phase transition to other BH with a positive heat capacity.
Finally, we investigate the  shadow bound of  favored region  for the BH  immersed in the DMH by comparing EHT observations.
\end{abstract}
\vspace{5mm}

\vspace{1.5cm}

\hspace{11.5cm}{Typeset Using \LaTeX}
\newpage
\renewcommand{\thefootnote}{\arabic{footnote}}
\setcounter{footnote}{0}

%%%% Introduction %%%%

\section{Introduction}
It is known that supermassive black holes founded at the center of galaxies have played an important role in galaxy formation and galaxy evolution.
The ground breaking  results of the Event horizon Telescope (EHT) collaboration have shed bright  light on  a new era of black hole (BH) observations.

Shadows and strong gravitational lensing was discussed in~\cite{Cunha:2018acu} and analytic study on the shadow including Ref.~\cite{Gralla:2019xty}  was reviewed in~\cite{Perlick:2021aok}.
 In addition, the shadows of  magnetically charged black holes from non-linear electrodynamics~\cite{Allahyari:2019jqz}, rotating regular black holes~\cite{Abdujabbarov:2016hnw}, non-rotating Kerr black hole~\cite{Atamurotov:2013sca},  BHs in the presence of plasma~\cite{Atamurotov:2015nra,Perlick:2015vta},  BH surrounded by dark matter~\cite{Konoplya:2019sns}, and BHs and naked singularities~\cite{Shaikh:2018lcc}   were studied.

Especially, the images of the M87* BH~\cite{EventHorizonTelescope:2019dse,EventHorizonTelescope:2019ths,EventHorizonTelescope:2019ggy} have inspired many studies on the  BH shadow  to test modified gravity theories.
Recently, the EHT results have focussed on   the center of our galaxy and revealed  interesting images of the  SgrA* BH~\cite{EventHorizonTelescope:2022wkp,EventHorizonTelescope:2022wok,EventHorizonTelescope:2022xqj}.
In this direction, the shadow of BH with scalar hair was used to test the EHT results~\cite{Khodadi:2020jij}, while the shadows of other BHs, worm holes, and naked singularities found from modified gravity theories have been employed to constrain their parameters~\cite{Vagnozzi:2022moj}.

It is  believed that astrophysical  BHs are not  isolated objects in the universe.
In addition to their accretion disks, they are immersed in the dark matter halo (DMH) which engulfs the whole galaxy~\cite{Bertone:2018krk}.
The authors of \cite{Cardoso:2021wlq} have suggested   an interesting  model of how  to embed a BH into a DMH by introducing  a Hernquist-type density  distribution that are observed in bulges and elliptical galaxies~\cite{Hernquist:1990be}. That is, a combined geometry of the BH with the DMH came as a solution to the Einstein theory coupled to an anisotropic fluid with $T_{\mu}~^\nu={\rm diag}(-\rho,0,p_t,p_t)$ where the mass (energy) density $\rho$
includes the Hernquist-type distribution.
 A lot of recent  studies on this and related models are found in~\cite{Konoplya:2021ube,Zhang:2021bdr,Jusufi:2022jxu,Konoplya:2022hbl,Figueiredo:2023gas,Zhao:2023itk,Stelea:2023yqo,Mollicone:2024lxy,Pezzella:2024tkf,Patra:2025mnj,Ovgun:2025bol}.
 Most of related works have included the computation of quasinormal modes for BHs surrounded by DMH, whereas  the effects of dark matter on the shadow of BHs were discussed  recently in~\cite{Xavier:2023exm,Myung:2024tkz,Macedo:2024qky}. Further, it is necessary  to understand how this combined geometry differs from others including a BH.
 It is suggested that the effect of dark matter on the shadow of a BH is similar to the effect of a constant scalar hair existing outside the event horizon on the shadow of a BH~\cite{Khodadi:2020jij,Myung:2024pob}, even though their equations of state are different.

In this work, we will study  a key feature, thermodynamics,  and  shadow bound of a BH with mass $M_b$ immersed in the DMH with mass $M$ and galaxy length-scale $a_0$. Here, we may consider an astrophysically relevant regime of  $M_b\ll M \ll a_0$~\cite{Navarro:1995iw}.
A key feature of the solution appeared in~\cite{Cardoso:2021wlq} is the presence of the redshift factor $e^{\Upsilon(r)}$ in the lapse function $f(r)$ when comparing with other black hole solutions.
We clarify that it has arisen from the mass distribution of the DMH  existing outside the event horizon because $\Upsilon(r)$ is zero for $M=0$.
Computing  the negative heat capacity of the BH surrounded by DMH indicates that  the BH could not be in thermal equilibrium with the DMH in
any regions outside the event horizon, contradicting to the Schwarzschild-AdS BH where AdS spacetime plays the role of  a confining box enclosing Schwarzschild BH~\cite{Prestidge:1999uq,Myung:2013uka}.
The first law of thermodynamics and Smarr formula for the BH favors the Newtonian ($1/a_0$) approximation but they disfavor the $1/a_0^2$ approximation obtained from  making large $a_0$-approximation.
The Newtonian Helmholtz free energy is used to show that there is no phase transition to other BH with positive heat capacity surrounded by a DMH.
We analyze and discuss the shadow radius (=critical impact parameter) by choosing three critical impact parameters.
We use the M87$^*$ and SgrA$^*$ shadow data obtained by the EHT collaboration to constrain  two parameters ($M,a_0$) of the DMH.

\section{BH surrounded by a DMH}

Firstly, we  introduce  the Hernquist-type density  distribution to describe a DMH solely as ~\cite{Hernquist:1990be}
\begin{equation}
\rho_{\rm H}(r)=\frac{Ma_0}{2\pi r(r+a_0)^3},
\label{H-den}
\end{equation}
where $M$ denotes the total  mass of a DMH and $a_0$ represents  a typical length-scale of the galaxy.
An exactly solution for a BH surrounded by a DMH is found through the Einstein cluster approach with  an anisotropic  stress-energy tensor
\begin{equation}
T_\mu~^\nu={\rm diag}(-\rho,0,p_t,p_t).
\end{equation}
Here,   one notes  a vanishing radial pressure $p_r=0$ to ensure a DMH.
Including a BH with mass $M_b$ at the center, the resulting spacetime is described by~\cite{Cardoso:2021wlq}
\begin{equation}
ds^2=-f(r)dt^2+\frac{dr^2}{1-\frac{2m(r)}{r}}+r^2d\Omega^2, \label{sss}
\end{equation}
where the lapse function $f(r)$ could be derived from the mass function
\begin{equation}
m(r)=M_b+\frac{Mr^2}{(r+a_0)^2}\Big(1-\frac{2M_b}{r}\Big)^2.
\end{equation}
\begin{figure*}[t!]
   \centering
   \includegraphics[width=0.6\textwidth]{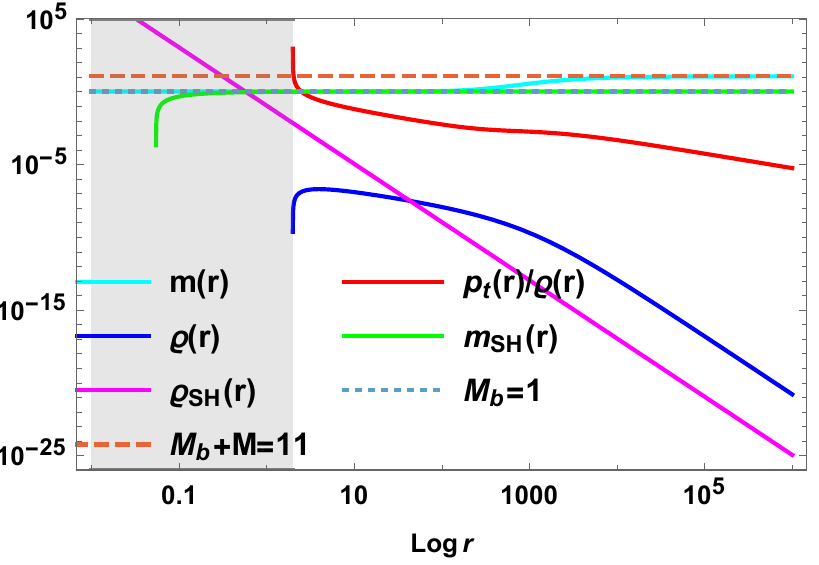}
\caption{ Mass function $m(r,M_b=1,M=10,a_0=1000)$, mass density $\rho(r,1,10,1000)$, $p_t(r,1,10,1000)/\rho(r,1,10,1000)$, and mass function  $m_{SH}(r,\tilde{m}=1,Q=0.1,S=0.1)$ for scalar hairy BH. These all are  loglog-function and shadow region denotes inside the event horizon at $r=r_+(=2)$.
We find a negative mass (energy) density ($\rho<0$) inside the event horizon, implying that the weak energy condition of $\rho\ge 0$ is violated.   }
\end{figure*}
We do not know how the gravitational collapse forms a BH at the center of the galaxy. It is known that $m(r)$ includes  a BH with mass $M_b$  in the aftermath of a BH-collapse at the center of the galaxy.
As is shown in Fig. 1, this mass function (cyan curve)  describes a BH with mass $M_b=1$ at small scales (even for $r=0$) and  it goes over to $M_b+M=11$  at large scales.
To understand how this mass function is unique, we introduce the mass function and energy density for the scalar hairy BH with BH mass $\tilde{m}$, electric  charge $Q$, and  scalar charge $S$ obtained from the Einstein-Maxwell-conformally coupled scalar theory as~\cite{Astorino:2013sfa,Myung:2024pob}
\begin{equation}
m_{SH}(r)=\tilde{m}-\frac{Q^2+S}{2r},\quad \rho_{SH}=-p_{r,SH}=\frac{Q^2+S}{r^4},\quad \phi=\sqrt{\frac{6}{\kappa}}\sqrt{\frac{S}{Q^2+S}}, \label{SH-m}
\end{equation}
where the last represents a constant scalar hair existing outside the event horizon. From Fig. 1, we observe that $m_{SH}(r)$ is zero at $r=0.055$ and 0.97 near the horizon ($r=\tilde{r}_+=1.94$) for $\tilde{m}=1,Q=0.1,S=0.1$. It is found that  $m_{SH}(r)= m(r)\simeq 1$ for $r\in[2,100]$.
However, $m_{SH}(r)$ differs from $m(r)$  inside the event horizon and
at large scales because any DMH is absent here.  This implies that  the transition of  $m(r)$ from $M_b=1$ at $r=100$  to $M_b+M=11$ at $r=10^{5}$ is due to the presence of  DMH and it will make the redshift factor in the lapse function.
In this case, the mass (energy) density $\rho$ is recovered by solving the continuity equation ($\rho=m'(r)/4\pi r^2$)   and  the tangential pressure $p_t$ are obtained from the Bianchi identity [$p_t=m \rho/2(r-2m)$] as
\begin{eqnarray}
\rho(r)&=&\frac{M(2M_b+a_0)}{2\pi r(a_0+r)^3}\Big(1-\frac{2M_b}{r}\Big), \label{rho-1} \\
p_t(r)&=& \frac{M(2M_b+a_0)(a_0^2 M_b+4M M_b^2+2a_0 M_b r-4M M_b r+M r^2+M_b r^2)}{4\pi r^2(r+a_0)^3(a_0^2+4M M_b+2a r-2M r+r^2)},                                                                       \label{pre-1}
\end{eqnarray}
where one finds from Fig. 1 that for $M_b=1,M=10,a_0=1000$,  $\rho= 0$ and $p_t=M/16\pi M_b(2M_b+a_0)^2=2\times 10^{-7}$ at  the event horizon ($r=2M_b$), while  $\rho\sim \rho_{\rm H}\simeq 10^{-21}$ and $p_{t}\simeq 10^{-27}$ at $r=10^6$.
We note that $\rho_{SH}(r,Q=0.1,S=0.1)(=p_{t,SH})$ has a similar behavior to $\rho(r,1,10,1000)$ outside the event horizon. Also, one notes that $p_{t,SH}/\rho_{SH}=1$ likes as $M_b=1$ for any regions.
Importantly, $p_t/\rho$ blows up at the event horizon, which means that the dominant energy condition ($\rho\ge p_t$) is violated near the event horizon.  Despite this violation, the region near the horizon contains almost
no  dark matter. The  mass density and  tangential  pressure  become arbitrarily  small in this area [$\rho(r=10)\simeq 10^{-7},p_t(r=10)\simeq 10^{-8}$] so that  they do not alter the spacetime formed by the BH. We note that $p_t/\rho$ crosses $p_{t,SH}/\rho_{SH}=1$ at $r=2.5$ and it decreases as $r$ increases.  We find a negative mass (energy)  density ($\rho<0$) inside the event horizon, which may imply that the weak energy condition of $\rho\ge 0$ is violated. However, this arises from putting  a BH at the center.  Actually, $\rho$ and $p_t$ are necessary to represent DMH (environment) existing outside the event horizon.

On the other hand, solving the Einstein equation of $G_{rr}=0$ together with an asymptotically flat spacetime
\begin{equation}
\frac{rf'}{2f}=\frac{m(r)}{r-2m(r)},
\end{equation}
the lapse function $f(r)$ is obtained analytically  as
\begin{eqnarray}
f(r)=\Big(1-\frac{2M_b}{r}\Big)e^{\Upsilon(r)} \label{shift-f}
\end{eqnarray}
with
\begin{equation}
\Upsilon(r)=\sqrt{M/\xi}\Big[-\pi +2\tan^{-1}\Big(\frac{r+a_0-M}{\sqrt{M\xi}}\Big)\Big], \quad \xi=2a_0-M+4M_b.
\end{equation}
Here, $e^{\Upsilon(r)}$ denotes  a redshift factor and its presence  indicates a key feature of the BH surrounded by a DMH. It ($\Upsilon(r)$) is 1(0) for $M=0$, implying that the redshift factor arises from the presence of  DMH distributed over outside the event horizon.
We note that the event horizon is located at $r_+=2M_b$ as for the Schwarzschild solution and the  ADM mass of spacetime  takes the form of $M_b+M$.
This solution could be regarded as a model for describing  a supermassive BH located at the center of a galaxy surrounded by a DMH.
To mimic observations of galaxies, one requires that $a_0 \geq10^4M$~\cite{Navarro:1995iw} and a hierarchy of scales: $M_{\rm bh} \ll M \ll a_0$. The compactness of DMH is measured by a quantity of ${\cal C}=\frac{M}{a_0}$ with $G=c=1$ unites and we consider the case of low-compactness with ${\cal C}\ll1$. It is worth noting that the redshift factor $e^{\Upsilon(r)}$ takes the form of $1-\frac{2M}{a_0}$ close the event horizon when making large $a_0$-approximation, while it is nearly 1 at a large distance.  Here,  $1-\frac{2M}{a_0}$ is regarded as the Newtonian redshift factor.
\begin{figure*}[t!]
   \centering
   \includegraphics[width=0.6\textwidth]{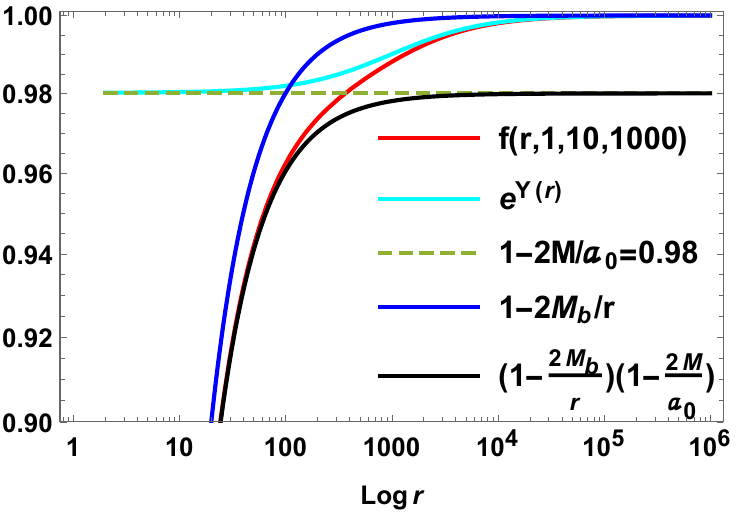}
\caption{ Lapse function $f(r,M_b=1,M=10,a_0=1000)$, its redshift factor $e^{\Upsilon(r)}$, and Schwarzschild form $(1-2M_b/r)$ outside the event horizon. These all are  loglog-function. We find that $f(r) \simeq (1-2M/a)(1-2M_b/r)$ near the event horizon,
 while $f(r) \simeq (1-2M_b/r)$ for $r\ge 10^{5}$. The redshift factor $e^{\Upsilon(r)}$ starts with  $1-2M/a_0=0.98$ at $r=2$  and ends with 1 at $r=10^6$.  }
\end{figure*}

Observing  Fig. 2, one finds that $f (r)$ is replaced by $(1-\frac{2M}{a_0})(1-\frac{2M_b}{r})$ near the event horizon,
while it is described by  the Schwarzschild metric $(1-\frac{2M_b}{r})$ for $r\ge 10^5$, ensuring consistency with Einstein gravity in vacuum. Here, it is important to note that the redshift factor $e^{\Upsilon(r)}$ (cyan curve) reflects the presence of mass function $m(r)$ for DMH existing outside the event horizon.  At this stage, we wish to clarify that  the role of this factor delays the arrival to an  asymptotically flat spacetime because of $(1-2M_b/r)|_{M_b=1,r=100}=0.98$ and $f(r=100,1,10,1000)=0.96$.
 Also, it propagates a further investigation of this BH in the next section. The combined geometry of a BH with DMH is described by the lapse function $f(r)$  in Eq.(\ref{shift-f}), which is a product of BH ($1-2M_b/r$) at the center and its environment (DMH, $e^{\Upsilon(r)}$)~\cite{Ovgun:2025bol}.

\section{Thermodynamic analysis}
As far as we know, there is no thermodynamic study on the BH surrounded by a DMH.
 To study thermodynamics of the BH surrounded by  a DMH, we need to define the Hawking temperature at $r=r_+=2M_b$ from the Unruh temperature experienced  by an static observer located outside the event horizon~\cite{Konoplya:2021ube}.
 It is given by
 \begin{equation}
 T_H(M_b,M,a_0)=\frac{1}{4\pi}\frac{f'(r)}{\sqrt{f(r)(1-\frac{2m(r)}{r})^{-1}}}\mid_{r=2M_b}=\frac{e^{\Upsilon(2M_b)/2}}{8\pi M_b}
 \end{equation}
 with
 \begin{equation}
 \Upsilon(2M_b)= \sqrt{M/\xi}\Big[-\pi +2\tan^{-1}\Big(\frac{2M_b+a_0-M}{\sqrt{M\xi}}\Big)\Big].
 \end{equation}
 Its Newtonian ($1/a_0$-approximation) temperature  $T^{N}_H$ and $1/a_0^2$ approximation temperature  $T^{a^{-2}_0}_H$ are given by
 \begin{equation}
 T^{N}_H=\frac{1}{8\pi M_b}\Big(1-\frac{M}{a_0}\Big), \quad T^{a^{-2}_0}_H=\frac{1}{8\pi M_b}\Big(1-\frac{M}{a_0}+\frac{M^2+12M_bM}{6a^2_0}\Big)
 \end{equation}
 when making large $a_0$-approximation to the redshift factor $e^{\Upsilon(r_+)/2}$. They ($T_H,T_H^N,T^{a^{-2}_0}_H$) all  are the same decreasing  functions of $M_b$ for ${\cal C} \ll1$, implying that there is no chance to possess a positive heat capacity.

 Making use of the first law of thermodynamics ($dM_b=T_HdS$), we obtain  two approximate  entropies  as
 \begin{eqnarray}
&& S^N(M_b,M,a_0)=\frac{4\pi M_b^2}{1-\frac{M}{a_0}}=\frac{A}{4}\Big(1+\frac{M}{a_0}\Big), \label{entropy-1} \\
 && S^{a^{-2}_0}=\frac{\pi a_0^2(12M_b M-(6a_0^2-6a_0M +M^2)\log[12M_bM+6a_0^2-6a_0M+M^2])}{3M^2}. \label{entropy-2}
 \end{eqnarray}
 In the Newtonian approximation, we find that the Smarr formula $M_b=2T_H^N S^N$ is well-defined.  On the other hand, the Smarr formula of $M_b=2T_H^{a^{-2}_0} S^{a^{-2}_0}$  is not satisfied in the $1/a_0^2$ approximation.
 Also, $S^{a^{-2}_0}$ is always  negative for ${\cal C} \ll1$, implying that it is unacceptable.
 This means  that the $1/a_0^2$ approximation is not a promising one to describe the thermodynamic aspects of a BH surrounded by a DMH.

 Furthermore, to test the local thermodynamic stability, one has to compute the heat capacity $C=(\partial T_H/\partial M_b)^{-1}$. It is fair to say that the thermal  stability (instability) can be achieved when $C>0(C<0)$.
 Its exact form is given by
 \begin{equation}
 C(M_b,M,a_0)=-\frac{8\pi M_b^2 e^{-\Upsilon(2M_b)/2}}{1 -\frac{2M_b M}{(2M_b+a_0)\xi}-\frac{M_b \sqrt{M}(\pi -2 \tan^{-1}[(2M_b+a_0-M)/\sqrt{M\xi}])}{\xi^{3/2}}},
 \end{equation}
  which leads to $-8\pi M_b^2$ for that  of  Schwarzschild BH  in the limit of $M\to 0$.
  Its approximated  heat capacities are derived from its approximated temperatures as
  \begin{equation}
  C^N=-8\pi M_b^2\Big(1+\frac{M}{a_0}\Big), \quad C^{a^{-2}_0}=-8\pi M_b^2\Big(1+\frac{M}{a_0}+\frac{5M^2}{6a_0^2}\Big)\label{ahc}
  \end{equation}
  which are the same forms obtained from  making large $a_0$-approximation on $C(M_b,M,a_0)$.
\begin{figure*}[t!]
   \centering
  \includegraphics[width=0.4\textwidth]{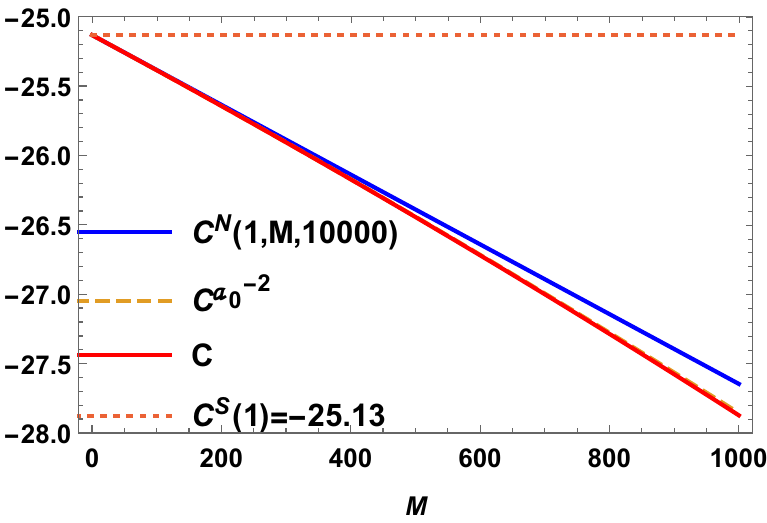}
 \hfill%
    \includegraphics[width=0.4\textwidth]{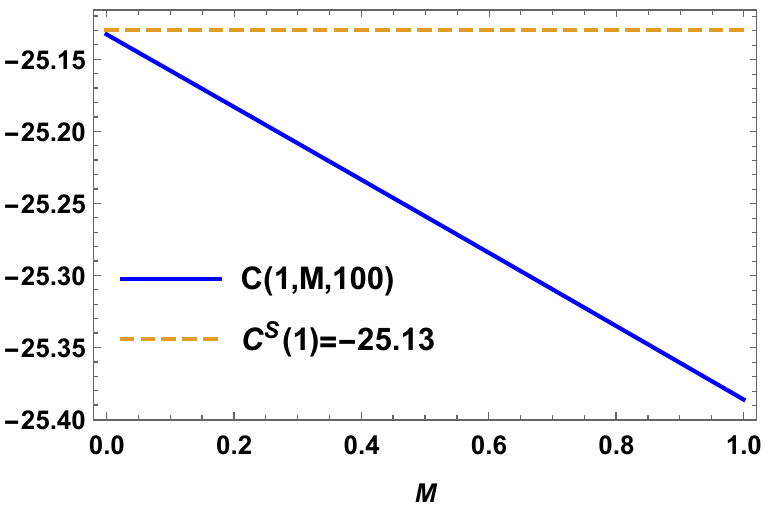}
\caption{(Left) Three heat capacities  $C^N(M_b=1,M,a_0=10000),~C^{-a_0^2}(1,M,10000),$ and $C(1,M,10000)$ as functions of DMH mass $M\in[1,1000]$.  They are nearly the negatively same  decreasing function. Here, $C^S(M_b)=-8\pi M_b^2$ represents the heat capacity for Schwarzschild BH with $M_b=1$. (Right) Enlarged heat capacity  $C(1,M,100)$ as a function of DMH mass  $M\in[0,1]$. It is  still a negatively  decreasing function.}
\end{figure*}
It is worthy to note that three of $C^N(M_b,10,10000),~C^{-a_0^2},$ and $C$ as functions of BH mass  $M_b$  are the negatively same  decreasing function.
As is shown in (Left) Fig. 3, three heat capacities are  the negatively same decreasing functions of DMH mass $M$, which means that the BH cannot be in a thermal equilibrium with whole galaxy because their heat capacities all are negative. This is similar to Schwarzschild BH in the asymptotically flat spacetime, whose heat capacity is given by $-8\pi M_b^2$. It is well known that  a BH with negative heat capacity is thermodynamically unstable.
The Schwarzschild BH looks like  a hot object in the asymptotically  flat spacetime (without heat reservoir). The AdS spacetime may play a role of the reservoir for the Schwarzschild  BH.

 From (Right) Fig. 3, we observe that the thermodynamic equilibrium of BH with DMH is not established in a relatively small region near the BH because the heat capacity $C(M_b=1,M,100)$ is always negative for the mass of DMH $M\in[0,M_b=1]$.  This implies that BH could be not in thermal equilibrium with the mass ($M$) of  DMH in
any regions, contradicting to the Schwarzschild-AdS black hole where AdS spacetime plays the role of a confining box enclosing the Schwarzschild BH~\cite{Prestidge:1999uq,Myung:2013uka}.
This is because a BH having  negative heat capacity in the asymptotically flat spacetime  is thermodynamically unstable.
 The thermodynamic influence of the environment (DMH) is limited to being small on the BH and it  does not alter the nature of BH  having  negative heat capacity.

On the other hand, the heat capacity of the Schwarzschild-AdS BH  becomes positive through Davies point (blow-up point), implying that Schwarzschild BH could be in thermal equilibrium with the negative cosmological constant.
That is, small AdS BH is unstable because of its negative heat capacity, while  large AdS BH is stable because of its positive  heat capacity. 
In addition, the heat capacity of the scalar hairy BH can be positive through Davies point, implying that this BH could be in thermal equilibrium with the scalar charge $S$~\cite{Astorino:2013sfa,Myung:2024pob}.
Interestingly, the heat capacity of Bardeen regular  BH can be positive through Davies point, suggesting that the BH could be in thermal equilibrium with the magnetic charge $g$~\cite{Ayon-Beato:2000mjt,Rodrigues:2018bdc,Quevedo:2024fga}.
It is worth noting that the equation of state for the last two cases is given by $\rho=-p_r$, whereas the equation of state for the DMH takes the form of $\rho\not=0,~p_r=0$ as is shown in Eqs.(\ref{rho-1})-(\ref{pre-1}).
In our case, we may  need a cavity or AdS spacetime  enclosing the combined geometry to get   a BH with positive heat capacity.

Finally, the global stability and phase transition are determined by the Helmholtz free energy~\cite{Myung:2007qt,Touati:2022zbm}. If it is positive (negative), it is globally unstable (stable). If the slope of Helmholtz free energy as a function of the Hawking temperature changes from negative to positive, there may be a phase transition. Here, we find  the  Newtonian free energy given by
\begin{equation}
F^N(M_b)=M_b-T^N_HS^N_{BH}=\frac{M_b}{2}\to F^N(T^N_H,M,a_0)=\frac{1}{16\pi T^N_H}\Big(1-\frac{M}{a_0}\Big).
\end{equation}
 \begin{figure*}[t!]
   \centering
  \includegraphics[width=0.4\textwidth]{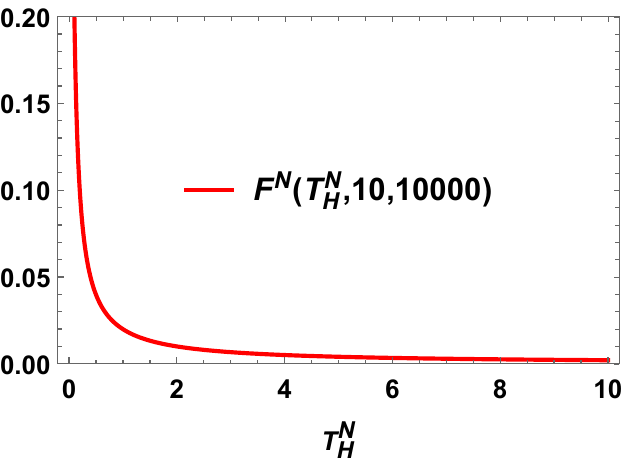}
 \hfill%
    \includegraphics[width=0.4\textwidth]{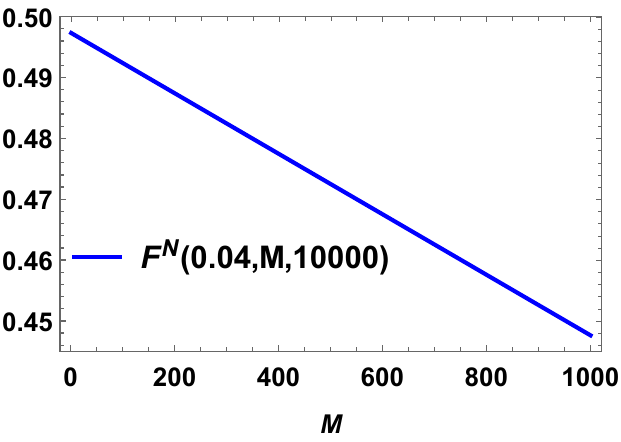}
\caption{(Left) Helmholtz free energy   $F^N(T^N_H,10,10000)$ as a function of  the Hawking temperature $T^N_H\in[0.1,10]$.  It is a monotonically decreasing function. (Right)
Helmholtz free energy   $F^N(0.04,M,10000)$ as an decreasing function of  the DMH mass $M\in[1,1000]$.}
\end{figure*}
As is shown in Fig. 4, the free energy is a monotonically  decreasing function of the Hawking  temperature  $T^N_H$ as well as it is a linearly decreasing function of the DMH mass $M$.
There is no change in the slopes of two curves, implying that there is no phase transition to other BH surrounded by a DMH.
In addition, we note that $F^N(M_b)=M_b/2$  is always  a  positive function of the BH mass $M_b$, ensuring  that it is impossible to achieve a global stability when varying $M_b$.

\section{Shadow bounds}
One has to  know the light ring radius $r_{\rm LR}$ to derive the shadow radius.
For this purpose, we introduce  the effective potential for null geodesics in  an equatorial plane~\cite{Xavier:2023exm}
\begin{equation}
V(r)=\frac{f(r)}{r^2}.
\end{equation}
A light ring  corresponds to a critical point of $V(r)$, that is,
\begin{equation}
V'(r)|_{r=r_{\rm LR}}=0. \label{pot-d}
\end{equation}
On the other hand, one obtains  an exact light ring radius  by finding  a real root to $r=3m(r)$~\cite{Cardoso:2021wlq}
\begin{eqnarray}
r_{\rm RL}(M_b,M,a_0)&=&M+M_b-\frac{2a_0}{3} \nonumber  \\
&-&\frac{2^{1/3}\zeta}{3(\eta+\sqrt{4(\zeta^3+\eta^2)})^{1/3}+\frac{1}{3\cdot 2^{1/3}}(\eta+\sqrt{4(\zeta^3+\eta^2)})^{1/3}} \label{photon-rad}
\end{eqnarray}
with
\begin{eqnarray}
\zeta&=&-a_0^2+12a_0M-9(M^2+M_b^2)-6a_0 M_b+18M M_b, \label{xi-p} \\
\eta&=& 2a_0^3+45a_0^2M-108a_0M^2+54(M^3+a_0M_b^2+M_b^3) \nonumber \\
&+&18a_0^2M_b+162 (a_0MM_b- M^2M_b). \label{eta-p}
\end{eqnarray}
For $M_b\ll M\ll a_0$, however,  one  obtains an approximate light ring
\begin{eqnarray}
r_{\rm{RL},a}=3M_b\Big(1+\frac{M_b M}{a_0^2}\Big) \label{gam-1}
\end{eqnarray}
when making large $a_0$-approximation on $r_{\rm RL}(M_b,M,a_0)$~\cite{Cardoso:2021wlq,Xavier:2023exm}. This means that there is no correction to $r_{\rm RL}$ up to  ${\cal O}(1/a_0)$ (in the Newtonian approximation).
\begin{figure*}[t!]
   \centering
   \includegraphics[width=0.4\textwidth]{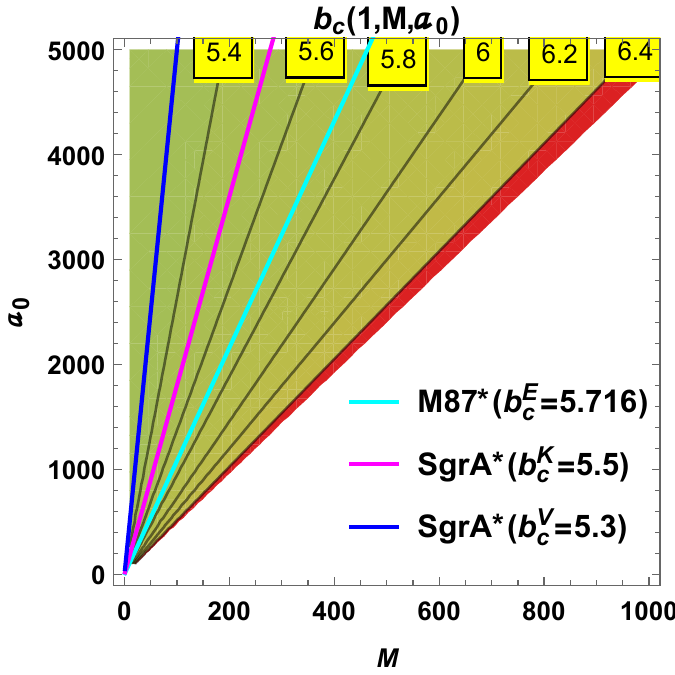}
\caption{Plot of the exact critical impact parameter $b_c(M_b=1,M,a_0)$ as functions of $M\in[1,1000]$ and $a_0\in[0,5000]$ with
 $M_b=1$. One has the lower limit of $b_c(1,0.001,5000)=b_c^S=5.196$.  Different black lines denote the lines for the same critical impact parameters.  Based on the EHT results, the favored (consistent with observations) region ($5.196\le b_c\le 5.716$) is separated from the disfavored (inconsistent with observations) region ($b_c>5.716$)  by the cyan line ($b_c^E=5.716$). This line could be represented precisely by $a_0(M)=10.8M$. For the SgrA$^*$, the magenta line ($b_c^K=5.5$) is given by $a_0(M)=18M$, while the blue line ($b_c^V=5.3$) is given by $a_0(M)=50M$. }
\end{figure*}

Now, we  compute the shadow radius to test the EHT and Keck results.
The shadow radius is defined by the critical impact parameter for the observer at infinity
\begin{equation}
r_{\rm sh}\to b_c=\frac{1}{\sqrt{V(r)}}
\mid_{r=r_{\rm RL}}=\frac{r}{\sqrt{f(r)}}\mid_{r=r_{\rm RL}}.\label{r-cri}
\end{equation}
Three types of critical impact parameters are available as
\begin{eqnarray}
b_{c}(M_b,M,a_0)&=&3\sqrt{3} M_be^{-\Upsilon(r_{\rm RL})/2}, \label{bc} \\
b_{c}^N&=&3\sqrt{3} M_b\Big(1+\frac{M}{a_0}\Big), \label{Nbc} \\
b_c^{a_0^{-2}}&=&3\sqrt{3} M_b\Big[1+\frac{M}{a_0}+\frac{5M^2-18M_bM}{6a_0^2}\Big], \label{abc}
\end{eqnarray}
where the last  is obtained by making large $a_0$-approximation on $b_c$~\cite{Macedo:2024qky}. The middle corresponds to the Newtonian approximation which is consistent with thermodynamic analysis.
This approximation came from $e^{\Upsilon(r)}=1-2M/a_0$ in the redshift factor when closing to the central BH. In this approximation, there is no correction to  the light ring as is shown in Eq.(\ref{gam-1}). 
Here, the DMH distribution ($M,a_0$) outside the BH determines  a modification of the redshift factor and  critical impact parameter solely, as was found in  Newtonian boson star profiles~\cite{Liebling:2012fv,Annulli:2020lyc}. We find from Eq.(\ref{abc})  that the leading order correction is due to gravitational redshift, but there are subdominant contributions whose detection is challenging if $M/a_0\le  10^{- 4}$.  It shows that the galactic content can affect  the shadow of BH   by terms of order $M^2/a_0^2 \le 10^{-8}$.

We check that $\lim_{M\to 0}b_c=3\sqrt{3}M_b (=b_c^S$, Schwarzschild case).
In Fig. 5, we show the  favored (disfavored) region by making use of the results  of the EHT collaboration for M87$^*$~\cite{EventHorizonTelescope:2019dse} and for SgrA$^*$~\cite{EventHorizonTelescope:2022xqj}. Here, we have to use the exact critical impact parameter $b_c$ only.
The lower limit  is found to be  $b_c^L(1,0.001,5000)= b_c^S=5.196$.
 If a relative deviation from the Schwarzschild  result ($b_c^S$) is less (greater) than 10\%, the solution is in the favored (disfavored) region~\cite{Xavier:2023exm}. The line of $b_c^{E}=5.716$ represents the upper limit for the favored region and it is also given by $a_0(M)=10.8M~({\cal C}=1/10.8)$.  Therefore, the favored region for shadow bound is represented by $a_0\ge10.8M$. For the SgrA*~\cite{EventHorizonTelescope:2022xqj}, the favored region is represented by $a_0\ge 18M(b_c^K=5.5,~{\cal C}=1/18)$ for Keck Observatory and it is small as  $a_0\ge 50M(b_c^V=5.3,~{\cal C}=1/50)$ for the Very Large Telescope Interferometer.
 However,  we remind the reader  that the observation of galaxies corresponds to the regime of $a_0\ge 10^4M$~\cite{Navarro:1995iw}.

\begin{figure*}[t!]
   \centering
   \includegraphics[width=0.4\textwidth]{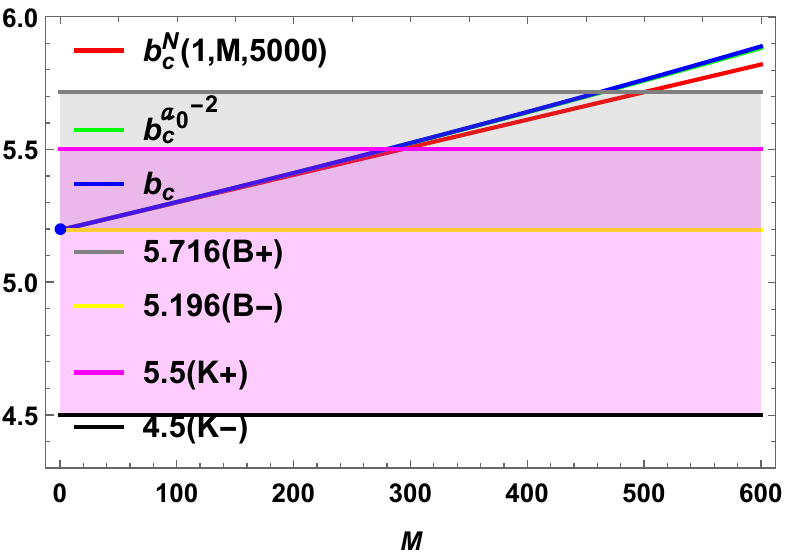}
 \hfill%
    \includegraphics[width=0.4\textwidth]{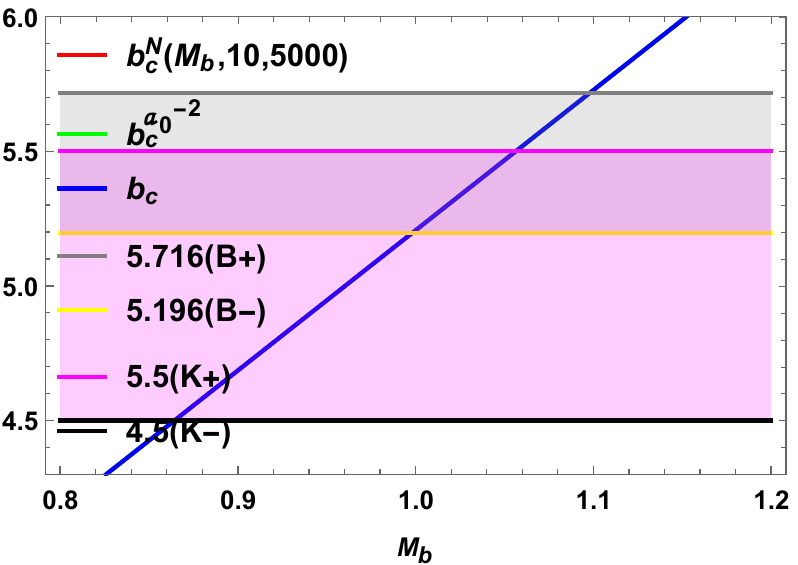}
\caption{(Left) Three critical impact parameters  of $b_c^N(M_b=1,M,a_0$=5000), $b_c^{a_0^{-2}}(1,M,5000),~b_c(1,M,5000)$ as  functions of DMH mass $M\in[0,600]$.  (Right) Three impact parameters of $b_c^N(M_b,M=10,a_0=5000),$ $b_c^{a_0^{-2}}(M_b,10,5000),~b_c(M_b,10,5000)$ are functions of BH mass $M_b\in[0.8,1.2]$.  }
\end{figure*}

Furthermore, we need to introduce  the  favored region when adopting two approximated critical impact parameters $b_c^N$ and $b_c^{a_0^{-2}}$.  If the relative deviation from the Schwarzschild  result is less  than 10\%, the solution is in the favorable  region with [5.196(E$-$),5.716(E+)] for the EHT observations (M87$^*$).   Also, we introduce the favored region [4.5(K$-$),5.5(K+)] for Keck Observatory of SgrA$^*$. As is shown in (Left) Fig. 6, three impact parameters $b_c^N(1,M,5000),$ $b_c^{a_0^{-2}}(1,M,5000),b_c(1,M,5000)$
belong to the favored region starting from $M=0$ to $M=500,465,462$ for M87$^*$ and $292,280,279$ for SgrA$^*$, respectively.  Here $b_c^N$ selects the largest $M$.
From (Right) Fig. 5, one finds that all of $b_c^N(M_b,10,5000),b_c^{a_0^{-2}},b_c$ are in the favored region
from $M_b=0.999$ to 1.098 for  M87$^*$ and from $M_b=0.864$ to 1.056 for  SgrA$^*$.

\section{Discussions}

We have investigated   thermodynamics and   shadow bound of  the BH  immersed  in the DMH.
In this case, we wish to point out that   the redshift factor $e^{\Upsilon(r)}$ has  captured  a key feature of the DMH  surrounding the BH, while the BH is still  described by Schwarzschild BH. 
This  factor  has  arisen from the mass function of DMH  which describes the mass distribution  over  outside the event horizon.

First of all, studying the heat capacity of the BH surrounded by DMH  implied  that  the BH could  not be  in thermal equilibrium with the DMH in
any regions. This contradicts to the Schwarzschild-AdS BH~\cite{Prestidge:1999uq,Myung:2013uka}.
 This means that  the thermodynamic influence of the DMH  (environment) is limited to being  small on the BH.    Therefore,  it  does not alter the nature of the BH with negative heat capacity which implies that the BH is thermodynamically unstable.  This phenomena is similar to the  influence of the DMH  on  classical and quantum radiation around the BH~\cite{Konoplya:2021ube}. 
 One may  introduce either   cavity or AdS spacetime enclosing the combined geometry to obtain  a thermodynamically stable BH.

In addition, it is desirable to note that the Newtonian ($1/a_0$) approximation provides a correct thermodynamic description for the BH surrounded by DMH  because the first law of thermodynamics and Smarr formula are satisfied. On the other hand, the $1/a^2_0$ approximation is  inappropriate for describing  its thermodynamics  because its Smarr formula is violated.
In this respect, we have employed the Newtonian Helmholtz free energy to show that there is no phase transition to other BHs with positive heat capacity.

We have used the M87* and SgrA$^*$ shadow data  to constrain  a relation between two parameters ($M,a_0$) of the DMH  surrounding  the central BH  with mass $M_b=1$.  
The shadow radius was analyzed  by the exact critical impact parameter $b_c$.
If a relative deviation from the Schwarzschild  result ($b_c^S=3\sqrt{3}M_b= 5.196$) is less (greater) than 10\%, the solution is in the favorable (disfavored) region~\cite{Xavier:2023exm}.
Here, the line of $b_c^{E}=5.716$ represents the upper limit and it corresponds to  $a_0(M)=10.8M(\mathcal{C}=1/10.8)$~\cite{Myung:2024tkz}. 
Therefore, the shadow bound for favored (consistent) region could be  represented by $a_0\ge10.8M$. 

It is desirable to  note that   the authors in~\cite{Macedo:2024qky}  have predicted  the compactness of DMH ($\mathcal{C}=1/12\sim 1/15$), while the authors~\cite{Xavier:2023exm} have proposed an interesting case of $\mathcal{C}=3/2$.
On the other hand, the shadow bound  is determined  by $a_0\ge 18M(b_c^K=5.5,~\mathcal{C}=1/18)$ for Keck Observatory and $a_0\ge 50M(b_c^V=5.3,~\mathcal{C}=1/50)$ for the Very Large Telescope Interferometer~\cite{EventHorizonTelescope:2022xqj}.
However, these $\mathcal{C}$  all are greater  than observations of galaxy [$a_0\ge 10^4M(b_c^V=5.3)\to\mathcal{C}=10^{-4}]$.  In the future, one has to explain this big difference if the present model is suitable for describing the BH surrounded by the DMH.

\vspace{1cm}

{\bf Acknowledgments} \\
 \vspace{1cm}

 This work was supported by the National Research Foundation of Korea(NRF) grant
 funded by the Korea government(MSIT) (NRF-2022R1A2C1002894).

\end{document}